%Paper: hep-ph/9310235
%From: HUANG@PIERRE.MIT.EDU
%Date: Wed, 6 Oct 1993 11:50:15 -0400 (EDT)

%
% Kerson Huang, "An Asymptotically Free Phi4 Theory"
% Plain TeX
% Figures available as hard copies. Send requests to
% Kerson Huang, 6-309, MIT, Cambridge MA 02139. huang@mitlns.
%
\baselineskip=24pt
\magnification=1200
\tolerance=10000
\nopagenumbers
\line{\hfil                                    CTP \#2208}
\line{\hfil                                    September, 1993}
\vskip 1truein
\centerline{An Asymptotically Free $\phi^4$ Theory}
\vskip 0.5truein
\centerline{Kerson Huang}
\vskip 0.5truein
\centerline{Center for Theoretical Physics, Laboratory for
Nuclear Science}
\centerline{       and Department of Physics}
\centerline{   Massachusetts Institute of Technology}
\centerline{   Cambridge, MA, USA 02139}
\bigskip
\centerline{                ABSTRACT}
The $\phi^4$ theory in $d=4-\epsilon$ dimensions has two
fixed points, which coincide in the limit $\epsilon\to 0$. One is a
Gaussian UV fixed point, and the other a non-trivial IR
fixed point.
They lead to two different continuum field theories.
The commonly adopted IR theory is ``trivial,''
behaves like perturbation theory,
and suggests an upper bound on the Higgs boson mass.
The UV theory is asymptotically free, and
does not impose a bound on the Higgs mass.
The UV continuum limit can also
be reached in d=4 with momentum or lattice cutoff.

\vfill
\eject
\pageno=1
\footline={\hss\tenrm\folio\hss} %restore page numbering

Renormalizable field theories generally fall into two categories:
infrared (IR)-stable, such as QED, or ultraviolet (UV)-stable, such as QCD.
They are based on fixed points with positive and negative
$\beta$-functions, respectively. The way to approach the continuum
limit is different for these cases. An IR fixed point
marks the end of a critical line
along which the correlation length diverges.
To attain the continuum limit, one can
approach any point on the critical line.
A UV fixed point, on the other hand, is not associated with a critical line,
and the continuum limit is obtained by approaching the  fixed point itself.

The $\phi^4$ scalar field theory in $d=4$
space-time dimensions is usually taken to be an IR theory,
because conventional perturbation theory yields a positive $\beta$-function.
We know from Wilson's $\epsilon$-expansion [1], however,
that the relevant fixed point is actually
a confluence of two fixed points, which are resolved in $d=4-\epsilon$
dimensions: a Gaussian UV fixed point, and a non-trivial IR fixed point
a distance $O(\epsilon)$ away. It is thus
not surprising that the single fixed point, into which they collapse
when $\epsilon\to 0$, has both IR and UV aspects. The IR aspect,
which dominates conventional thinking,
leads to a ``trivial'' theory, the approach to which turns out to be
logarithmically slow. For practical
purposes, therefore, the cutoff can be kept large but finite, yielding
an effective interacting theory.
Requirements of self-consistency [2], however, imposes a bound
on the boson mass, (which we shall refer to as the Higgs mass.)

The alternative continuum limit base on the UV aspect of the fixed point
leads to an asymptotically free theory, which allows for
broken symmetry, but places no bound on the Higgs mass.
The UV theory is trivial as well, at least in
perturbation theory. It thus appears
to be a truly free field with possible broken symmetry.
To contrast the IR and UV continuum limits, we first review
the usual IR theory.

For simplicity, we
consider a one-component scalar field $\phi$, with bare Lagrangian density
$$
L={1\over 2}(\partial\phi)^2
-{1\over 2}m^2_0 \phi^2 - {1\over 4}\lambda_0 \phi^4 \eqno(1)
$$
We are interested in the case of broken symmetry, with
$m^2_0 < 0$.
The usual perturbation theory in d=4 gives to
lowest order $\beta(\lambda)=c\lambda^2$, with $c=3/(64\pi^2)$.
This leads to a renormalized coupling $\lambda_r$ given by
$$
{1\over\lambda_r}={1\over\lambda_0}+c\ln{\Lambda\over m} \eqno(2)
$$
where $\Lambda$ is the cutoff, and
 $m$ a physical mass. Since $\lambda_0 > 0$, we must
have $(1/\lambda_r) > c\ln(\Lambda/m)$. Thus, the continuum limit
is ``trivial,'' though the limit is approached with logarithmic slowness:
$$
\lambda_r<{1\over c\ln(\Lambda/m)}\mathop{\to 0}_{\Lambda\to\infty} \eqno(3)
$$
The insensitivity to the cutoff enables us to maintain a nonzero $\lambda_r$
by keeping the cutoff large but finite. In doing this, however,
consistency dictates that $m<\Lambda$.
For $m^2_0<0$, the tree-level vacuum field is given by
$\langle\phi\rangle = \sqrt{-3m^2_0/\lambda_0}$. Assuming
a similar relation between renormalized quantities,
${m/\langle\phi_r\rangle}\propto\sqrt{\lambda_r}$, and
using (2) and the experimental value
$\langle\phi_r\rangle\approx 250$ GeV,
Dashen and Neuberger estimates $m <$ 900 GeV [2].

The scenario above can be clarified by looking at the fixed-point
structure of the RG equations in $d=4-\epsilon$. Introducing the
dimensionless couplings
$$\eqalign{
r_0&=m_0^2\Lambda^{-2}\cr
u_0&=\lambda\Lambda^{-\epsilon}\cr
}\eqno(4)
$$
we have [1]
$$\eqalign{
b {du\over d b}&=\epsilon u-Au^2\cr
b {dr\over d b} &= B r+C r^2-D ru\cr } \eqno(5)
$$
where $A,B,C,D$ are positive constants. These equations describe the
evolution of the coupling constants under
coarse-graining, whereby $\Lambda$ is effectively
decreased to $\Lambda/b$. Note that $b$ increases towards the infrared.

The RG trajectories are schematically depicted in Fig.1 for $\epsilon>0$.
There are two fixed points near the origin in the $r$-$u$ plane: a Gaussian
UV fixed point at the origin, and a non-trivial IR fixed point
a distance $O(\epsilon)$ away.  The RG trajectory that
flows into the IR fixed point is the critical line $r=r_c(u)$,
which divides the plane into regions corresponding to the
symmetric and ``broken'' phases of the theory, and along which the
correlation length is infinite.
We can approach the continuum limit by
following a path in the parameter space that intersects
the critical line. The continuum theory lies
either in the symmetric or broken phase, depending on the direction
of approach. For d=4, the scaling behavior in the neighborhood
of the critical line is governed by the mean-field critical
exponents $\beta=\nu=1/2$, with logarithmic corrections [3]:
$$\eqalign{
m &\sim t^{1/2}(\ln t)^{-1/6}\cr
\langle\phi\rangle &\sim t^{1/2}(\ln t)^{1/3}\cr } \eqno(6)
$$
where $t=|r-r_c|$. Thus, $m/\langle\phi\rangle\sim \ln t^{-1/2}$.
Assuming $u\sim 1/t$, we have $u_r\sim 1/(\ln t)$, and thus
$m/\langle\phi\rangle\sim \sqrt{u_r}$.
This reproduces the naive perturbative behavior (4).

The above scenario, which is based on perturbation theory, has been
confirmed in Mont-Carlo simulations [5]--[9].
The existence of a phase with broken symmetry in this limit
was demonstrated in Ref.[7]. The logarithmic scaling corrections
have been verified in Ref.[8], and the bound on the Higgs mass
lowered to about 600 GeV [9].
{}From all evidence, we can conclude that there is a low-energy
effective theory,
which looks like perturbation theory, and which is not
very sensitive to the cutoff. The single important experimental
consequence is the bound on the Higgs mass.

The existence of an UV fixed point for $d=4-\epsilon$
enables us to define an alternative continuum field theory, which will
be asymtotically free, since the UV fixed point is at $u_0=0$.
To approach this limit, we should choose a path in the broken phase that
avoids the critical line, as illustrated by $P$ in Fig.1.
The point $B$ denotes the bare system at some cutoff $\Lambda$,
and $R$ denotes the renormalized system. The procedure is to hold
$R$ fixed, and choose a sequence of
bare systems $B$  receding along $P$ into the fixed point, which
corresponds to $\Lambda=\infty$. The limit $\epsilon\to 0$ is
taken afterwards.

A quantity of physical interest is the
dimensionless ratio $m/\langle\phi_r\rangle$, whose limit at the
Gaussian fixed point can be calculated. For
a Gaussian fixed point, scaling is governed by the
mean-field exponents $\beta=\nu=1/2$, with no logarithmic corrections.
Thus,
$$
m/\langle\phi_r\rangle\to {\rm Constant}  \eqno(7)
$$
with no restriction on the mass $m$. The procedure is familiar
from lattice QCD, where
one computes some dimensionless quantity, {\it e.g.},
the string tension in units of the mass gap, in the limit
as the bare coupling goes to zero [9].

The exponent $\beta$ for the Gaussian fixed point is sometimes
reported incorrectly as $(d-2)/4$, which would have given
$m/\langle\phi_r\rangle\to 0$ for any $\epsilon>0$. The correct value is
$\beta=1/2$, which reflects the fact that $u$ is
a ``dangerous irrelevant variable,'' in that it is needed to stabilize
the system in the broken phase [10].

To solve (5), let us put $b=-\ln(k/k_0)$, where
$k$ is the momentum scale at the renormalization point, and
$k_0$ an arbitrary reference scale. Integrating the first equation
yields the running coupling constant
$$
u(k)\propto\epsilon (k/k_0)^{-\epsilon} \eqno(8)
$$
which exhibits asymptotic
freedom for any $\epsilon>0$. Since it vanishes linearly with
$\epsilon$, the theory is also ``trivial,'' at least according to
perturbation theory.
Non-perturbative effects, however, may render the theory non-trivial.
Since the RG trajectory flows away from the Gaussian fixed point,
perturbation theory will eventually break down, and we cannot
rule out the possibility that the coupling will grow at lower energies.

For $\epsilon>0$, the separation of the IR and UV fixed points make
it clear that there exists an UV continuum limit distinct from the
IR limit. Can we reach the UV continuum limit in in d=4,
when the two fixed points coincide?
The answer is yes, for the fixed point has both IR and UV manifestations.
It acts as an UV fixed point for the trajectory running away from it,
i.e., the horizontal axis in Fig.2. The trouble is that the negative
axis, which lies in the broken phase, is unphysical. We have
to approach it through a sequence of trajectories.

Along a trajectory passing close to the origin in Fig.2,
points upstream are attracted towards an approximate IR fixed point,
while those downstream are repelled by an approximate
UV fixed point. Near the fixed point, the $r$-$u$ plane
can be divided into IR-like and UV-like regions by a fuzzy crossover line.
To approach the UV continuum limit, we can choose a sequence of bare
systems that go into the fixed point along some path $C$ that
lies in the UV region, ortogonal to the critical line.

Our considerations can be generalized to the case of a complex
Higgs field $\phi$. With the transformation
 $\phi = \sqrt{\rho}\exp(i\theta)$,
where $\rho$ and $\theta$ are canonically conjugate,
the Lagrangian density assumes the form
$$
L={1\over 2}\rho (\partial\theta)^2
+{1\over 4\rho}(\partial\rho)^2 - {m^2_0\over 2}\rho^2
- {\lambda_0\over 4}\rho^4						\eqno(9)
$$
In the continuum limit, $\rho$ should become a free field with broken symmetry,
while vortex activity remains possible in the $\theta$ sector.
An earlier study [11] of vortex rings in the  Weinberg-Salam model
suggests that there are unstable macroscopic rings related to
``sphalerons'' [12], and weakly-decaying microscopic ones with mass in
excess of 2.5 TeV. These excitations can exist in both the IR and UV
theories; but in the UV theory they would be the only excitations.

A proposal by Consoli and Stevenson [13] may be related to the UV limit
discussed here. They discussed a procedure based on the
one-loop effective potential that yields a negative $\beta$-function,
and the value $m/\langle\phi_r\rangle=\sqrt{8}\pi$, which puts the
Higgs mass at about 2 TeV.

There seems little doubt that the Higgs
field is but an effective field, useful in a  certain energy domain.
The question is whether it is a low-energy effective theory or
a high-energy one. Conventional thinking regards it
as a low-energy theory. In this  view, the Higgs boson is a composite
of more elementary constituents, $t\bar t$ quark pairs, for example.
Our proposal presents an alternative, which views
the Higgs as just as tightly bound as
quarks and leptons. At present there does not seem to be experimental
evidence favoring one view over another.
The UV theory would be favored if, for example,
a Higgs boson is not found, and no new physics emerges,
by the time experiments reach an energy scale of 600 GeV.

I thank Mehran Kardar, Patrick Lee, and Yue Shen for useful conversations.
This work is supported in part by funds provided by the U.S.
Department of Energy under contract \# DE-AC02-76ER03069.
\bigskip
\centerline{\bf References}
\bigskip
\item{[1]} See K.G. Wilson and J. Kogut, Phys. Rep. {\bf C12} (1974) 75;
E. Br\'ezin, J.C. Le Guillou, and J. Zinn-Justin in
{\it Phase Transitions and Critical Phenomena}, vol. 6, ed.
C. Domb and M.S. Green (Academic Press, London, 1976).
\medskip
\item{[2]} R. Dashen and H. Neuberger, Phys. Rev. Lett., {\bf 50} (1983) 1897.
\medskip
\item{[3]} A.I. Larkin and D.E. Khumel'nitskii, Sov. Phys. JETP {\bf 29}
(1969) 1123.
\medskip
\item{[4]} B. Freedman, P. Smolinsky, and D. Weingarten, Phys. Lett.,
{\bf 113B} (1982) 481.
\medskip
\item{[5]} For reviews see C.B. Lang, Nucl. Phys,, {\bf B265} (1986) 630;
P. Hasenfratz in {\it Lattice 88, Proceedings of the 1988 Symposium
on Lattice Field Theory}, ed. A.S. Kronfeld and P.B. Mackenzie
(North-Holland, Amsterdam 1989).
\medskip
\item{[6]} K. Huang, E. Manousakis, and J. Polonyi, Phys. Rev., {\bf 35}
(1987) 3187.
\medskip
\item{[7]} J. Kuti and Y. Shen, Phys. Rev. Lett., {\bf 60} (1988) 85.
There appears to be a disagreement between this and
Ref.[6] on the behavior of the wave-function renormalization constant.
\medskip
\item{[8]}  J. Kuti, L. Lin, and Y. Shen, Phys. Rev. Lett., {\bf 61}
(1988) 678.
\medskip
\item{[9]} See C. Rebbi, {\it Lattice Gauge Theories and Monte Carlo
Simulations} (World Scientific, Singapore, 1983).
\medskip
\item{[10]} N. Goldenfeld, {\it Lectures on Phase Transitions and
Renormalisation Group} (Addison-Wesley, Reading, MA, 1991), p.359.
\medskip
\item{[11]} K. Huang and R. Tipton, Phys. Rev., {\bf D23} (1981) 3050.
\medskip
\item{[12]} N.S. Manton, Phys. Rev., {\bf D28} (1983) 2019;
F.R. Klinkhamer and N.S. Manton, Phys. Rev., {\bf D30} (984) 2212.
\medskip
\item{[13]} M. Consoli and P. Stevenson, Rice University preprint (1993).
\vfill
\eject
\centerline{\bf Figure Captions}
\noindent
Fig.1. RG trajectories for d=4-$\epsilon$ ($\epsilon\to 0^+$). Arrows
point toward the infrared.
The IR and UV fixed points define different continuum field theories.
The IR continuum limit can be reach by approaching the critical line.
The UV continuum limit is reached by choosing a sequence of bare systems $B$,
approaching the origin along $P$, with the renormalized system
$R$ held fixed.
\medskip
Fig.2. RG trajectories for d=4.
The UV continuum limit can be reached by choosing a sequence
of bare systems $B_1,B_2,\cdots$ along a path $C$
in the UV-like region, orhtogonal to the critical line.

\end